\documentclass[%
reprint,
 amsmath,amssymb,
 aps,
]{revtex4-2}

\usepackage{graphicx}
\usepackage{dcolumn}
\usepackage{bm}
\usepackage{amsmath}
\usepackage{nccmath}
\usepackage[dvipdfmx]{hyperref}
\usepackage[dvipdfmx]{xcolor}
\usepackage{mathtools}
\usepackage{natbib}
\usepackage{booktabs}
\usepackage{mathrsfs}
\usepackage{multirow}

\begin{document}

\preprint{APS/123-QED}

\title{Effects of neck and nuclear orientations on the mass drift in heavy ion collisions}

\author{Shota Amano$^{1}$, Yoshihiro Aritomo$^{1}$ and Masahisa Ohta$^{2}$\\
\scriptsize{$^{1}$Kindai University Higashi-Osaka, Osaka 577-8502, Japan}\\
\scriptsize{$^{2}$Konan University Kobe, Hyogo 658-8501, Japan}\\
\scriptsize{e-mail: 2144340401y@kindai.ac.jp}\\
}

\date{\today}

\begin{abstract}
\noindent{\bf Background:} We clarified that the fusion hindrance in heavy ion collisions is caused by the expansion of the neck bridge at the early stage of collision \lbrack\href{https://doi.org/10.1103/PhysRevC.108.014612}{\color{blue}Phys.\;Rev.\;C\;{\bf 108},\;014612\;(2023)}\rbrack, however, our discussion was limited to the trajectory analysis. For getting the reliable fusion cross section, it is important to understand the fusion process connecting with multinucleon transfer and also the process depending on the target orientation in detail. Especially, the effects of target orientation on the multinucleon transfer process have not been discussed so far in our model.\\
{\bf Purpose:} First, we investigate precisely the start time of the neck expansion relevant to the mass transfer. The main aims of this paper are to discuss the mass drift in the collision with the different target orientations within the dynamical approach based on the fluctuation-dissipation theorem in the reaction $^{32}$S + $^{232}$Th. \\
{\bf Method:} The orientation effects are incorporated within the framework of the Langevin equation with three nuclear deformation parameters as the degree of freedom and the two-center shell model (TCSM) for the potential energy of the system. \\
{\bf Results:} The start time of the neck expansion was presumed 10 zs (zs = $10^{-21}$ s) by analysis in the several entrance channels. By taking account of the target nuclear orientation, a strong mass-angle correlation was obtained which is compatible with the experimental data. There was a large difference in the mass transfer or the mass drift mode in the fusion process between the tip and the side collision. \\
{\bf Conclusions:} Not only ``delayed relaxation'' of the neck but also the nuclear orientation effects have an important role in the strong correlation between fragment mass and its emitting angle.
The mass evolution toward mass symmetry is slower than the standard mass drift mode assuming an exponential-type function. Particularly, the mass drift of the tip collision follows the slow mass drift mode assuming a Fermi-type function rather than an exponential-type function, which is related to the different features of the maximum neck cross-sectional area in the sticking process.

\end{abstract}

\maketitle


\section{Introduction}
To date, superheavy elements up to $Z=118$ have been identified. If the next new element is successfully synthesized, it will reach the 8th period of the periodic table. To determine how many elements exist, it is necessary to investigate the nuclear structure by producing nuclei reaching ``island of stability'' \cite{oganessian2015beachhead}.
However, in the conventional fusion reactions (hot fusion reactions \cite{RevModPhys.72.733,PhysRevC.74.044602,hofmann2007reaction} and cold fusion reactions \cite{hofmann1995production,armbruster1985production,munzenberg1988recent,doi:10.1143/JPSJ.76.043201}), the synthesized nucleus is limited to have enough neutrons to reach ``island of stability''. In order to overcome this problem, multinucleon transfer (MNT) reactions, which were widely used in the 1970s and 1980s, are attracting attention again.

The usefulness of MNT reactions for the studies in unexplored areas on the nuclear chart (unknown neutron-rich heavy and superheavy nuclei)  has been discussed \cite{PhysRevC.101.024610,zhu2022approach,zhu2019possibilities,heinz2022nucleosynthesis,PhysRevC.83.044618,Zagrebaev_2008,Zagrebaev_2013,PhysRevC.87.034608,PhysRevLett.101.122701} to reach “islands of stability”.
Recently, the advantage of MNT reactions in producing neutron-rich nuclei ($N=126$) in contrast to fragmentation reactions \cite{PhysRevC.89.024616,KURCEWICZ2012371,TAIEB2003413} was revealed \cite{PhysRevLett.115.172503}.

%
In MNT reactions, various nuclei with different excitation energies are produced over a wide angular range of separating fragments. The optimal conditions for producing the aimed nuclei using MNT reactions have been investigated \cite{heinz2022nucleosynthesis,PhysRevC.108.024602,zagrebaev2015production,PhysRevC.101.024610,PhysRevC.108.024602,PhysRevC.106.014606,PhysRevResearch.5.L022021,PhysRevC.99.014613,saiko2022multinucleon}.
It is also necessary to investigate the stability (fission barrier) of the produced nuclei in MNT reactions. The height of the fission barrier for nuclei produced by MNT reactions is  strongly related to the angular momentum induced by the transferred nucleons.
However, the determination of the induced angular momentum cannot be evaluated directly.
Therefore, the angular momentum of the produced nuclei is determined from the correlation of the other relevant observables \cite{Nishio2023,PhysRevC.105.L021602}.
In this point, the reliable theoretical method for determining the property of fragments produced by MNT is needed.

%
On the one hand, the fusion hindrance in heavy ion collisions due to the expansion of the neck bridge in the early stage of collision have been described and clarified in our previous paper \cite{amano2023dynamical}. Further investigations are needed to clarify the fusion dynamics for heavy and superheavy elements.
A correlation between fission fragment mass and scattering angle is also important for understanding the reaction dynamics of fusion.
The formation of heavy and superheavy elements is accomplished by progressing the mass equilibration during the reaction timescale. Typically, the timescales of the reaction process are $10$ zs to $10^{5}$ zs for fusion-fission (FF), $1$ zs to $10$ zs for quasifission (QF), and $\leq1$ zs for quasielastic (QE) and deep-inelastic collision (DIC). The time unit of zeptosecond (zs) is equal to $10^{-21}$ s. Both mass evolution and timescale of collision processes have been discussed by analyzing the mass angle distribution  (MAD) \cite{TOKE1985327,PhysRevC.36.115,du2013mapping,prasad2016exploring,du2011predominant,tanaka2021mass,tanaka2023competition}. From the features of the MAD, the dominant reaction process is understandable \cite{du2013mapping}.
The features of MAD are mainly distinguished into following three types: The first one is FF and the mass-symmetric fission is dominant. However, there is no correlation between mass and angle, because the memory of entrance channel disappears during a lot of rotations after two nuclei stick together.
The second type is predominant by QF which the mass evolution is suppressed due to the experience of less rotation in comparison with type 1. The most obvious difference from the type 1 is that there is a strong correlation between mass and angle. This correlation is very important to analyze the time (sticking time) from contact to scission in the QF reaction. To clarify the sticking time of QF leads to give important insights to the dynamics of the nuclear fusion process.
The third type of MAD accompanied with QE and DIC has the feature that the fission fragment mass yield populates close to the initial mass and the grazing angle.

In the 1980s, by using experimental MAD, a correlation between sticking time and mass transfer (mass drift)  was systematically analyzed in $^{238}$U induced reaction on various targets \cite{TOKE1985327,PhysRevC.36.115}. As a result, the universality of the mass drift curve has been reported.
Recently, in the ﬁrst trial for direct experimental measurement of the angular momentum dependence of ﬁssion fragments, the sticking time of fast quasiﬁssion (FQF) related to the angular momentum has been evaluated \cite{tanaka2021mass}. Due to the strong role of fluctuation in the process of the mass evolution of FQF for intermediate angular momentum, the mass evolution toward mass symmetry is slower than the standard mass drift curve \cite{TOKE1985327,PhysRevC.36.115}.
Besides, the results comparing the experimental MAD with the simulated MAD present the slower mass drift curve rather than the standard mass drift one in various mass evolution angles \cite{prasad2016exploring}.

%
%
In the following section, the Langevin-type approach taking into account the orientation effects is described.
The calculation results of MAD and $M_\text{R}$ depending on the start time of the neck expansion are shown in \ref{sec3A}, choosing several entrance channels. In addition, we discuss the importance of the neck expansion in the mass-angle correlation and simultaneously determine the appropriate value of the start time of the neck expansion.
The nuclear orientation effects for the mass drift in the $^{32}$S + $^{232}$Th reaction is followed in \ref{sec3B}, where the mass drift and the MAD are discussed in detail and the summary is noted in the final section.

\section{Theoretical framework}
\subsection{Transition to the adiabatic state}
We adopt the dynamical model based on the multidimensional Langevin equations, which similar to unified model \cite{Zagrebaev_2007}.
Early in the collision, the reaction stage of the nucleon transfer consists of two parts. First, at the approaching stage the system is placed in the ground state of the projectile and target because the reaction proceeds is too fast for nucleons to occupy the lowest single-particle levels. Next, the system relaxes to the ground state of the entire composite system which changes the potential energy surface to an adiabatic one.
Therefore, we treat the transition of two reaction stages with a time-dependent weighting function:
\begin{eqnarray}
&&V=V_\mathrm{{diab}}\left(q\right)f\left(t\right)+V_\mathrm{{adiab}}\left(q\right)\left[1-f\left(t\right)\right], \\
&&f\left(t\right)=\exp{\left(-\frac{t}{\tau_\mathrm{DA}}\right)}.
\label{pot}
\end{eqnarray}
Here, $q$ denotes a set of collective coordinates representing nuclear shape. The diabatic potential $V_\text{diab}\left(q\right)$ is calculated within the double-folding method with Migdal nucleon-nucleon forces \cite{Zagrebaev_2005,Zagrebaev_2007,zagrebaev2007potential,migdal1967theory}.
The adiabatic potential energy $V_\text{adiab}\left(q\right)$ of the system is calculated using an extended two-center shell model \cite{zagrebaev2007potential} and described the details in \ref{C}. As a characteristic of the diabatic potential, "potential wall" appears due to the overlap region of collision system which corresponds to the hard core representing the incompressibility of nuclear material.
$t$ is the interaction time and $f\left(t\right)$ is the weighting function included the relaxation time $\tau_\text{DA}$. We use the relaxation time $\tau_\text{DA}=0.1$ zs proposed in \cite{bertsch1978collision,cassing1983role,PhysRevC.69.021603}.
With the two-center parameterizations \cite{maruhn1972asymmetrie,sato1978microscopic}, the nuclear shape which represents by three deformation parameter is defined as follows:
$z_{0}$ (distance between the centers of two potentials),
$\delta$ (deformation of fragment), and $\alpha$ (mass asymmetry of colliding nuclei);
$\alpha=\frac{A_{2}-A_{1}}{A_{2}+A_{1}}$, where $A_{1}$ and $A_{2}$ not only stand for the mass numbers of the projectile and target respectively \cite{Zagrebaev_2005,ARITOMO20043} but also are then used to indicate mass numbers of the two fission (light and heavy) fragments.
The parameter $\delta$ is defined as $\delta=\frac{3\left(a-b\right)}{\left(2a+b\right)}$, where $a$ and $b$ represent the half  length of the ellipse axes in the $z_{0}$ and $\rho$ directions, respectively \cite{maruhn1972asymmetrie}.
In addition, we use scaling to save computation time and use the coordinate $z$ defined as $z=\frac{z_{0}}{\left(R_\text{CN}B\right)}$, where $R_\text{CN}$ denotes the radius of the spherical compound nucleus and the parameter $B$ is defined as $B=\frac{\left(3+\delta\right)}{\left(3-2\delta\right)}$.
We solve the dynamical equation numerically. Therefore, we restricted the number of degrees of freedom as three deformation parameters to avoid the huge calculation time.

\subsection{Consideration of the nuclear orientation}
\begin{figure}[t]
\centering
\includegraphics[keepaspectratio, width=\linewidth]{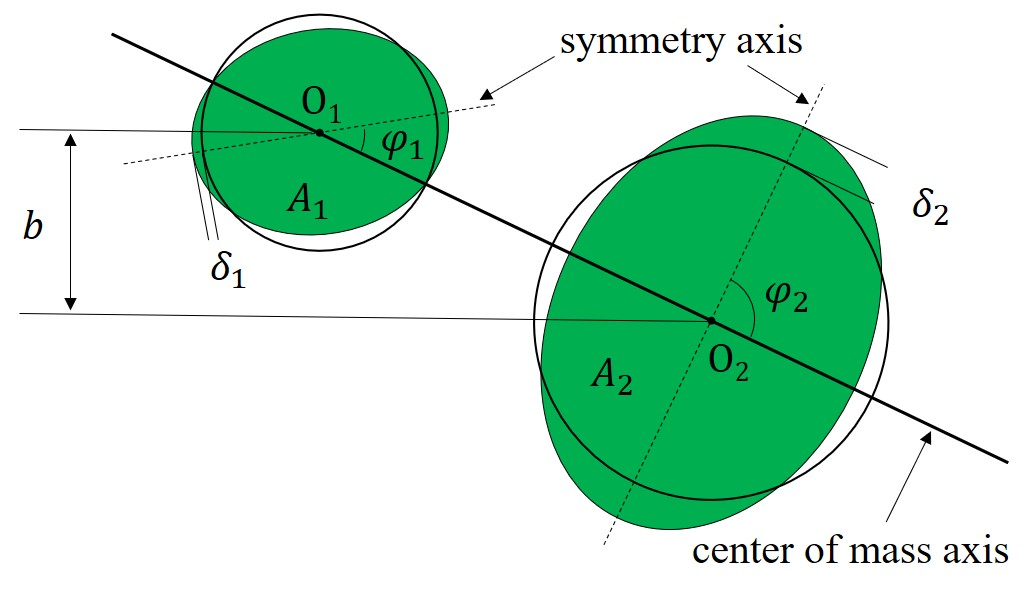}
\caption{Geometrical diagram when both the deformed incident nucleus and the target nucleus collide with each other in the different initial orientation angles.
}\label{fig_geo}
\end{figure}
In order to consider the orientation effects when using the deformed nucleus, we use the axially-symmetric potential $V_\text{ax}^\text{sym}$ the axially-asymmetric one $V_\text{ax}^\text{asym}$ in the stage of the diabatic states. When two deformed nuclei collide with each other, mainly there are four colliding patterns: the so-called tip-to-tip ($\varphi_{1}^{0}=\varphi_{2}^{0}=0$), side-to-side ($\varphi_{1}^{0}=\pi/2, \varphi_{2}^{0}=\pi/2$), tip-to-side ($\varphi_{1}^{0}=0, \varphi_{2}^{0}=\pi/2$) and side-to-tip ($\varphi_{1}^{0}=\pi/2, \varphi_{2}^{0}=0$). $\varphi_{1}^{0}$ and $\varphi_{2}^{0}$ denote the initial orientation angle of the projectile and target nuclei corresponding to $\varphi_{1}$ and $\varphi_{2}$ shown in Fig.~\ref{fig_geo}.
In this paper, we select the various $\varphi_{2}^{0}$ of the deformed $^{232}$Th nucleus fixing $\varphi_{1}^{0}=0$ of the spherical $^{32}$S nucleus when we investigate the collision system of $^{32}$S + $^{232}$Th.
In the case of the tip collision ($\varphi_{2}^{0}=0$), we use the diabatic potential of axially-symmetric states $V_\text{diab}=V_\text{ax}^\text{sym}$. If calculations for the side collision ($\varphi_{2}^{0}=\pi/2$) or colliding patterns of the others ($0<\varphi_{2}^{0}<\pi/2$) are performed, we use the diabatic potential of axially-asymmetric states $V_\text{diab}=V_\text{ax}^\text{asym}$. Here, note that $V_\text{ax}^\text{asym}$ depends on the nuclear orientation angle.
Besides, while $V_\text{diab}$ of the axially-asymmetric states transits to $V_\text{adiab}$ which is the axially-symmetric states, the ellipsoid deformations are adjusted with restores of the systems using the time-dependent form, and the final ellipsoid deformations $\delta_{i}^\text{fin}$ \cite{PhysRevC.99.014613} finished the restoration of the axially-symmetry is obtained as
\begin{eqnarray}
&&\tilde{\delta}=\delta_{i}f_\mathrm{res}\left(t\right)+\delta_{i}^\mathrm{fin}\left[1-f_\mathrm{res}\left(t\right)\right], \\
&&f_\mathrm{res}\left(t\right)=\exp{\left(-\frac{t}{\tau_\mathrm{res}}\right)}, \\
&&\delta_{i}^\mathrm{fin}=\left(1+\delta_{i}\right)\left[\delta_{i}\left(2+\delta_{i}\right)\mathrm{sin}^2\varphi_{i}^{0}+1\right]^{-\frac{3}{4}}-1.
\label{def_res}
\end{eqnarray}
$f_\text{res}$ is the weighting function including the relaxation time $\tau_\text{res}$, which performs the restoration from the axially-asymmetric system to the axially-symmetric one. We use $\tau_\text{res}=1$ zs in this paper. This value involved in the restoration of the system is also has been used in Ref. \cite{PhysRevC.99.014613}.
%
In the collision system using the deformed $^{232}$Th nucleus, we treat as $\delta_{i}\sim0.18$ with $\delta_{i}=3\beta_{2}/(\beta_{2}+\sqrt{16\pi/5})$ \cite{PhysRevC.85.044614}.
The quadrupole deformation $\beta_{2}$ for the deformed nucleus $^{232}$Th is 0.207 \cite{MOLLER1995185}.

\subsection{Adiabatic potential}
\label{C}

The neck parameter $\epsilon$ including in the two-center parameterizations is adjusted in Ref. \cite{YAMAJI1987487}. Reproduce the available data assuming different values between the entrance and exit channels of the reaction. In the present paper, we use $\epsilon = 1 $ for the entrance channel and $\epsilon = 0.35$ for the exit channel.
This treatment is used in Refs. \cite{zagrebaev2007potential,PhysRevC.85.044614}.
We assume the time dependence of the potential energy with the finite range liquid drop model, which is denoted by the characteristic relaxation time of the neck $t_{0}$ and the variance $\Delta_{\epsilon}$ as follows:

\begin{eqnarray}
&&V_\mathrm{{LDM}}\left(q,t\right)=V_\mathrm{{LDM}}\left(q,\epsilon=1\right) f_{\epsilon}\left(t\right) \nonumber \\
&& \qquad \qquad \qquad +V_\mathrm{{LDM}}\left(q,\epsilon=0.35\right) [1-f_{\epsilon}\left(t\right)],  \\
&&V_\mathrm{{LDM}}\left(q,\epsilon\right)=E_\mathrm{S}\left(q,\epsilon\right)+E_\mathrm{C}\left(q,\epsilon\right), \\
&&f_{\epsilon} = \frac{1}{1+\exp\left(\frac{t-t_{0}}{\Delta_{\epsilon}}\right)}, \label{fe}
\end{eqnarray}
where the symbols $E_\text{S}$ and $E_\text{C}$ stand for generalized surface energy and Coulomb energy, respectively \cite{PhysRevC.20.992}.
The temporal form of $\epsilon$ has been used commonly \cite{zagrebaev2007potential,saiko2022multinucleon,PhysRevC.99.014613,PhysRevC.96.024618}. We fixed $\Delta_{\epsilon}$ as $0.1$ zs used in the Ref. \cite{amano2023dynamical}.
If we use $t_{0}=0$ zs, $\epsilon$ parameter starts to relax from 1 to 0.35 as soon as two nuclei contact.

The adiabatic potential energy given a value of $\epsilon$ and a temperature of a system is defined as
\begin{eqnarray}
&&V_\mathrm{{adiab}}\left(q,t,L_\text{tot},T\right) \nonumber \\
&&\quad=V_\mathrm{{LDM}}\left(q,t\right)+V_\mathrm{SH}\left(q,T\right)+V_\mathrm{rot}\left(q,L_\mathrm{tot}\right).
\label{adipot}
\end{eqnarray}
$V_\text{SH}$ is the shell correction energy that takes into account the temperature dependence as
\begin{eqnarray}
&&V_\mathrm{SH}\left(q,T\right)=E_\mathrm{shell}^{0}\left(q\right)\Phi\left(T\right), \\
&&E_\mathrm{shell}^{0}\left(q\right)=\Delta E_\mathrm{shell}\left(q\right) + \Delta E_\mathrm{pair}\left(q\right),\\
&&\Phi\left(T\right)=\exp\left(-\frac{E^{\ast}}{E_{d}}\right),
\end{eqnarray}
where $E_\text{shell}^{0}$ indicates the microscopic energy at $T$ = 0, which is calculated as the sum of the shell correction energy $\Delta E_\text{shell}$ and the pairing correlation correction
energy $\Delta E_\text{pair}$. $T$ is the temperature of the compound nucleus calculated from the intrinsic energy of the composite system.
$\Delta E_\text{shell}$ is calculated by the Strutinsky method \cite{STRUTINSKY19681, RevModPhys.44.320} from the single-particle levels of the two-center shell model potential \cite{maruhn1972asymmetrie,suek74,10.1143/PTP.55.115} as the difference between the sum of single-particle energies of occupied states and the averaged quantity.
$\Delta E_\text{pair}$ is evaluated in the Bardeen-Cooper-Schrieffer (BCS) approximation as described in Refs. \cite{RevModPhys.44.320, NILSSON19691}. The averaged part of the pairing correlation energy is calculated assuming that the density of single-particle
states is constant over the pairing window. The pairing strength constant is related to the average gap parameter $\tilde{\Delta}$ by solving the gap equation in the same approximation and adopting $\tilde{\Delta} = 12/ \sqrt{A_\text{CN}}$ suggested in \cite{NILSSON19691} by considering the empirical results for the odd-even mass difference \cite{PhysRevC.90.054609}. $A_\text{CN}$ is the compound nucleus mass.
The temperature dependence factor $\Phi\left(T\right)$ is explained in Ref. \cite{ARITOMO20043}, where $E^{\ast}$ indicates the excitation energy of the compound nucleus. $E^{\ast}$ is given $E^{\ast}=a_\text{lev}T^{2}$, where $a_\text{lev}$ is the level density parameter. The shell damping energy $E_{d}$ is selected as 20 MeV. This value is given by Ignatyuk \textit{et~al}. \cite{ignatyuk1975phenomenological}.
The rotational energy $V_\text{rot}$ generated from the total angular momentum $L_\text{tot}$ represents as
\begin{eqnarray}
V_\text{rot}\left(q,L_\text{tot}\right) \qquad \qquad \qquad \qquad \qquad \qquad \qquad \qquad \qquad \nonumber \\
=\frac{\hbar^{2}L\left(L+1\right)}{2\mathscr{I}\left(q\right)}+\frac{\hbar^{2}L_{1}(L_{1}+1)}{2\Im_{1}(q)}+\frac{\hbar^{2}L_{2}(L_{2}+1)}{2\Im_{2}(q)}, \qquad
\end{eqnarray}
where $L$ and $L_{1,2}$ are the relative angular momentum (orbital angular momentum) and the angular momentum for the heavy and light fragments, respectively. $\mathscr{I}\left(q\right)$ is the rigid body moment of inertia with deformation, which is multiplied by two.


\subsection{Dynamical equations}
The trajectory calculations are performed on the time-dependent unified potential energy \cite{Zagrebaev_2005,Zagrebaev_2007,ARITOMO20043} using the multidimensional Langevin-type equations \cite{Zagrebaev_2005,ARITOMO20043,PhysRevC.80.064604} as follows:
%
\begin{eqnarray}
&&\frac{dq_{i}}{dt}=\left(m^{-1}\right)_{ij}p_{j},  \\
&&\frac{dp_{i}}{dt}=-\frac{\partial V}{\partial q_{i}}-\frac{1}{2}\frac{\partial}{\partial q_{i}}\left(m^{-1}\right)_{jk}p_{j}p_{k} \nonumber \\
&&\qquad \qquad \qquad -\gamma_{ij}\left(m^{-1}\right)_{jk}p_{k}+g_{ij}\Gamma_{j}\left(t\right), \\
&&\frac{d\vartheta}{dt}=\frac{\hbar L}{\mathscr{I}}, \label{rot}\\
&&\frac{d\varphi_{1}}{dt}=\frac{\hbar L_{1}}{\Im_{1}}, \\
&&\frac{d\varphi_{2}}{dt}=\frac{\hbar L_{2}}{\Im_{2}}, \\
&&\frac{dL}{dt}=-\frac{1}{\hbar}\frac{\partial V}{\partial\theta}-\gamma_\text{tan}\left(\frac{L}{\mu_{R}R^{2}}-\frac{L_{1}}{\Im_{1}}a_{1}-\frac{L_{2}}{\Im_{2}}a_{2}\right)R \nonumber \\
&&\qquad \qquad +\frac{R}{\hbar}\sqrt{\gamma_\text{tan}T}\Gamma_\text{tan}\left(t\right),  \\
&&\frac{dL_{1}}{dt}=-\frac{1}{\hbar}\frac{\partial V}{\partial\varphi_{1}}+\gamma_\text{tan}\left(\frac{L}{\mu_{R}R^{2}}-\frac{L_{1}}{\Im_{1}}a_{1}-\frac{L_{2}}{\Im_{2}}a_{2}\right)a_{1}\nonumber \\
&&\qquad \qquad -\frac{a_{1}}{\hbar}\sqrt{\gamma_\text{tan}T}\Gamma_\text{tan}\left(t\right),  \\
&&\frac{dL_{2}}{dt}=-\frac{1}{\hbar}\frac{\partial V}{\partial\varphi_{2}}+\gamma_\text{tan}\left(\frac{L}{\mu_{R}R^{2}}-\frac{L_{1}}{\Im_{1}}a_{1}-\frac{L_{2}}{\Im_{2}}a_{2}\right)a_{2} \nonumber \\
&&\qquad \qquad -\frac{a_{2}}{\hbar}\sqrt{\gamma_\text{tan}T}\Gamma_\text{tan}\left(t\right).
\end{eqnarray}
%
The collective coordinates $q_{i}$ represent $z, \delta$, and $\alpha,$ the symbol $p_{i}$ denotes momentum conjugated to $q_{i}$, and $V$ is the multidimensional potential energy.
$m_{ij}$ and $\gamma_{ij}$ stand for the shape-dependent collective inertia and friction tensors, respectively.
We adopted the hydrodynamical inertia tensor $m_{ij}$ in the Werner-Wheeler approximation for the velocity field \cite{PhysRevC.13.2385}.
The one-body friction tensors $\gamma_{ij}$ are evaluated within the wall-and-window formula \cite{RANDRUP1984105, PhysRevC.21.982}.
The normalized random force $\Gamma_{i}\left(t\right)$ is assumed to be white noise: $\langle \Gamma_{i} (t) \rangle$ = 0 and $\langle \Gamma_{i} (t_{1})\Gamma_{j} (t_{2})\rangle = 2 \delta_{ij}\delta (t_{1}-t_{2})$.
According to the Einstein relation, the strength of the random force $g_{ij}$ is given as $\gamma_{ij}T=\sum_{k}{g_{ik}g_{jk}}$.
$\vartheta$ and $\mu_{R}$ indicate the relative orientation of nuclei and the reduced mass.
$R$ is the distance between the nuclear centers $\text{O}_{1}\text{O}_{2}$ as shown in Fig.~\ref{fig_geo}.
$\varphi_{1}$ and $\varphi_{2}$ stand for the orientation angles of each nucleus (See Fig.~\ref{fig_geo}).
$a_{1,2}=\frac{R}{2}\pm\frac{R_{1}-R_{2}}{2}$ is the distance from the center of the fragment to the middle point between the nuclear surfaces, and $R_{1,2}$ is the nuclear radii.
The total angular momentum $L_\text{tot}=L+L_{1}+L_{2}$ is preserved.
The phenomenological nuclear friction forces for separated nuclei are expressed in terms of the tangential friction $\gamma_\text{tan}$ and the radial friction $\gamma_{R}$ using the Woods-Saxon radial form factor suggested in Ref. \cite{Zagrebaev_2005}. The treatment of $\gamma_\text{tan}$ and $\gamma_{R}$ are described in our previous papers \cite{PhysRevC.106.024610,amano2023dynamical}.

\subsection{Mass ratio $M_\text{R}$ and scattering angle $\theta_\text{c.m.}$}
\begin{figure}[t]
\centering
\includegraphics[keepaspectratio, width=\linewidth]{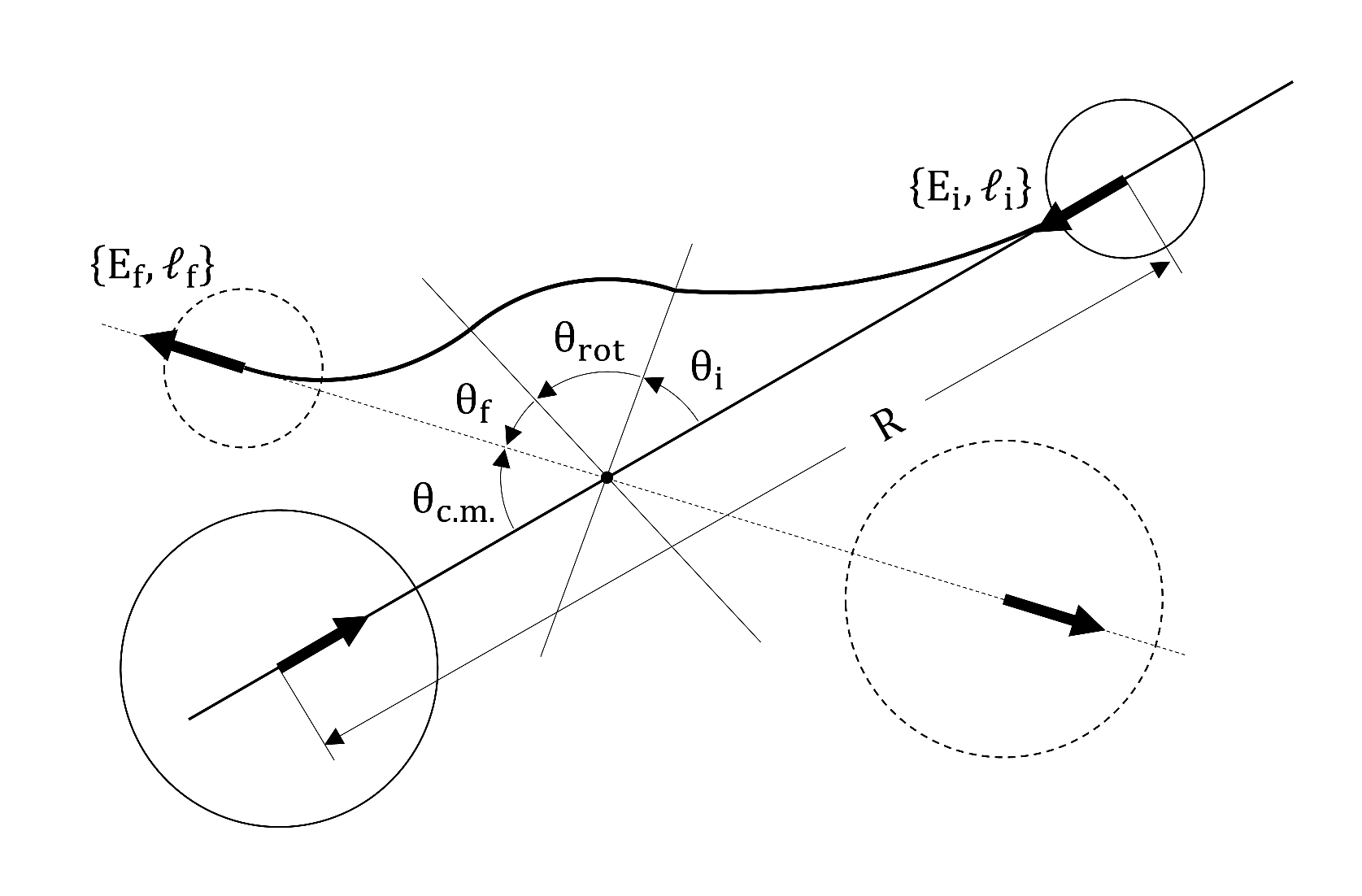}
\caption{Schematic determined the scattering angle $\theta_\text{c.m.}$ (in the center of mass system). The Coulomb scattering angles $\theta_i$ and $\theta_f$ determined by the Coulomb trajectories in entrance and exit channels. The angle $\theta_\text{rot}$ is the rotation angle of the system between contact and scission.
}\label{fig_theta}
\end{figure}
The MAD is presented by the mass-angle matrix using $M_\text{R}$ and the scattering angle $\theta_\text{c.m.}$.
The projectile-like mass ratio $M_\text{R}$ is determined as $M_R=A_1/A_\text{CN}$. Consequently, the target-like mass ratio is expressed as $M_R=1-M_R$.
Regarding the scattering angle $\theta_\text{c.m.}$ is determined from
\begin{eqnarray}
\theta_\text{c.m.}=\pi-\theta_{i}-\theta_\text{rot}-\theta_{f},
\end{eqnarray}
as shown schematically in Fig.~\ref{fig_theta}. $\theta_i$ and $\theta_f$ are the Coulomb scattering angle of initial and final trajectories. $\theta_\text{rot}$ is the rotation angle of the system between contact and scission, which is obtained from eq.~(\ref{rot}).
The Coulomb scattering angle of the incoming nuclei is obtained with the incident center-of-mass energy $E_{i}$ and each initial relative angular momentum (orbital angular momentum) $L_{i}$ as follows
\begin{eqnarray}
\theta_{i}=\text{arctan}\left(\sqrt{\eta_{i}^{c}}\right), \qquad
\eta_{i}^{c}=\frac{2E_{i}\hbar^{2} L_{i}^{2}}{\mu_{0}\alpha_{c}^2},
\end{eqnarray}
where $\mu_{0}$ is the initial reduced mass and $\alpha_{c}=Z_{p}Z_{t}e^{2}/4\pi\epsilon_0$. $Z_{p}$ and $Z_{t}$ are the charges of the projectile and target nuclei.
The Coulomb scattering angle $\theta_f$ of the outgoing nuclei is slightly complicated. $\theta_f$ is determined as follows
\begin{eqnarray}
\theta_{f}=&&\text{arctan}\left(\sqrt{\eta_{f}^{c}}\right)-\text{arctan}\left(\kappa \sqrt{\eta_{f}^{n}}\right), \label{theta_f} \\
\eta_{f}^{c}=&&\frac{2E_{f}\hbar^{2} L_{f}^{2}}{\mu_{0}\alpha_{c}^2}, \qquad
\eta_{f}^{n}=\frac{2E_{f}\hbar^{2} L_{f}^{2}}{\mathscr{I}R^{2}\alpha_{c}^2},
\end{eqnarray}
where $E_{f}$ and $L_{f}$ are the final energy after dissipating and the final relative angular momentum.
$\kappa$ is given by $\kappa=\frac{R}{\hbar^{2} L_{f}^{2}/(\mathscr{I}R^{2}\alpha_{c})+R}$.
At large angular momentum the second term in eq.~(\ref{theta_f}) is almost negligible. Thus, $\theta_{f}$ is determined by the value of $\eta_{f}^{c}$. Besides, $\theta_{f}$ is given by $\theta_{f} \simeq \theta_{i}$ because the final energy and the final relative angular momentum are almost unchanged from the incident energy and the initial angular momentum at large angular momentum.
Consequently, at large angular momentum we obtain the scattering angle function
\begin{eqnarray}
\theta_\text{c.m.}&=&\pi-(\theta_{i}+\theta_{f})=\pi-2\theta_{i} \nonumber \\
&=&\pi-\text{2arctan}\left(\sqrt{\frac{2E_{i}\hbar^{2} L_{i}^{2}}{\mu_{0}\alpha_{c}^2}}\right).
\end{eqnarray}
The scattering angle of the target-like nucleus is determined by $\pi-\theta_\text{c.m.}$.



\section{RESULTS AND DISCUSSION}
\label {sec3}

\subsection{Determination of the start time of the neck expansion
}
\label {sec3A}
\begin{figure*}[t]
\centering
\includegraphics[keepaspectratio, width=\linewidth]{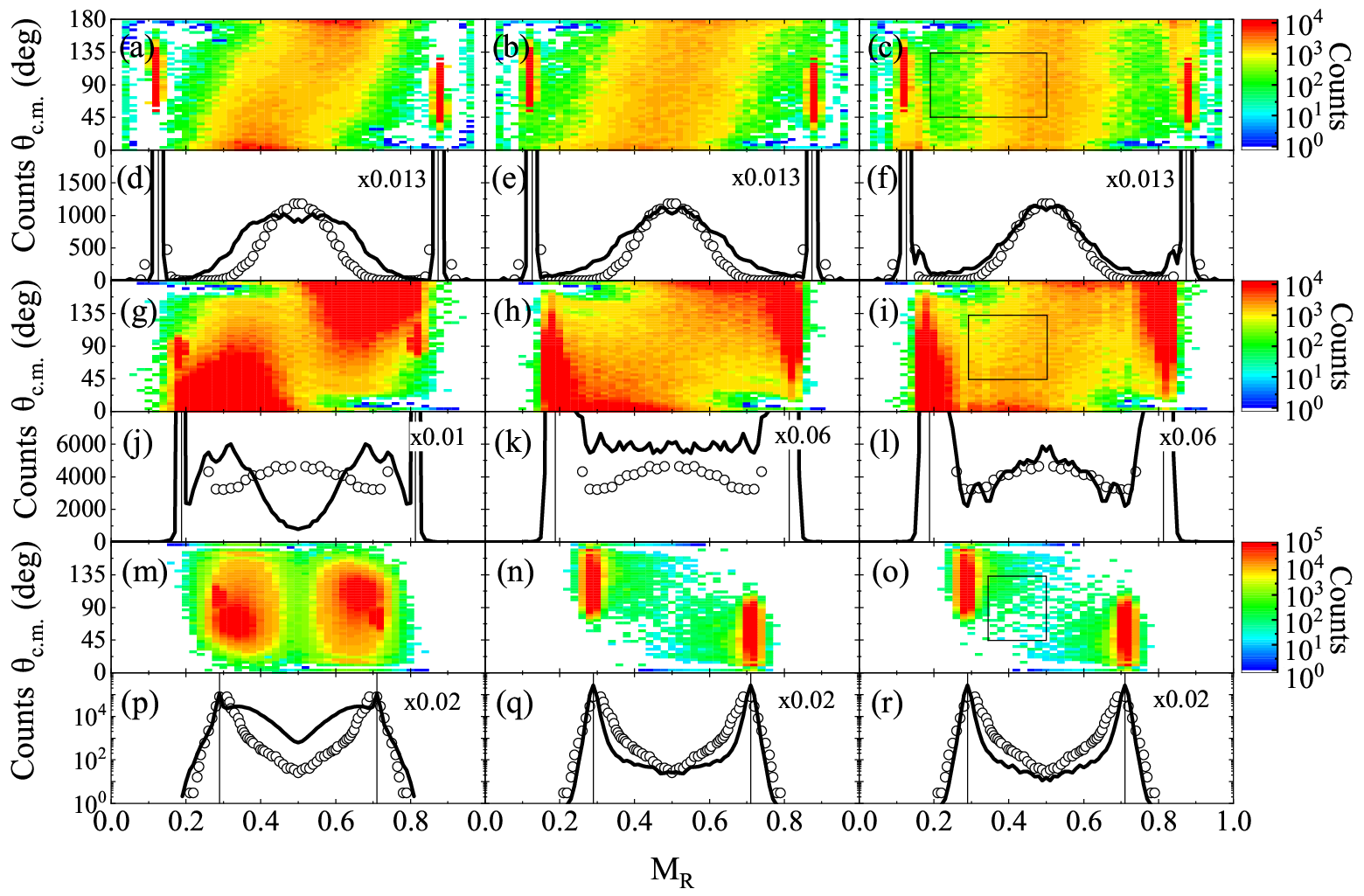}
\caption{Characteristic start time of the neck expansion t$_{0}$ dependence of the mass angle distributions (a)-(c), (g)-(i), (m)-(o) and $M_\text{R}$ distributions (d)-(f), (j)-(l), (p)-(r) for the different entrance channels. Panels (a)-(f), (g)-(l), and (o)-(r) are the collision systems of $^{30}$Si, $^{48}$Ti, and $^{86}$Kr with $^{208}$Pb at $E_\text{c.m.}/V_\text{bass}=$ 1.05, 1.173, and 1.084, respectively. The Bass barrier energies $V_\text{bass}$ for using $^{30}$Si, $^{48}$Ti, and $^{86}$Kr are 128.54 MeV, 195.18 MeV, and 302.63 MeV, respectively \cite{bass1980nuclear}. 
The calculations (black thick lines) used $t_{0}=0$ zs, $5$ zs, and $10$ zs in the left side, the middle side, and the right side panels, respectively. The theoretical value is multiplied by the scale factor on the right side of the $M_\text{R}$ distribution. The black vertical thin lines indicate the initial mass ratio $M_\text{R}^{0}$. 
The symbols are the experimental data \cite{williams2013evolution, du2013mapping, mg2002fusion}. 
The black rectangles in panels (c), (i), and (o) are $M_\text{R}$ and angular ranges, which are used for Fig. \ref{fig5} (see text).
}\label{fig3}
\end{figure*}
\begin{figure*}[t]
\includegraphics[scale=0.6]{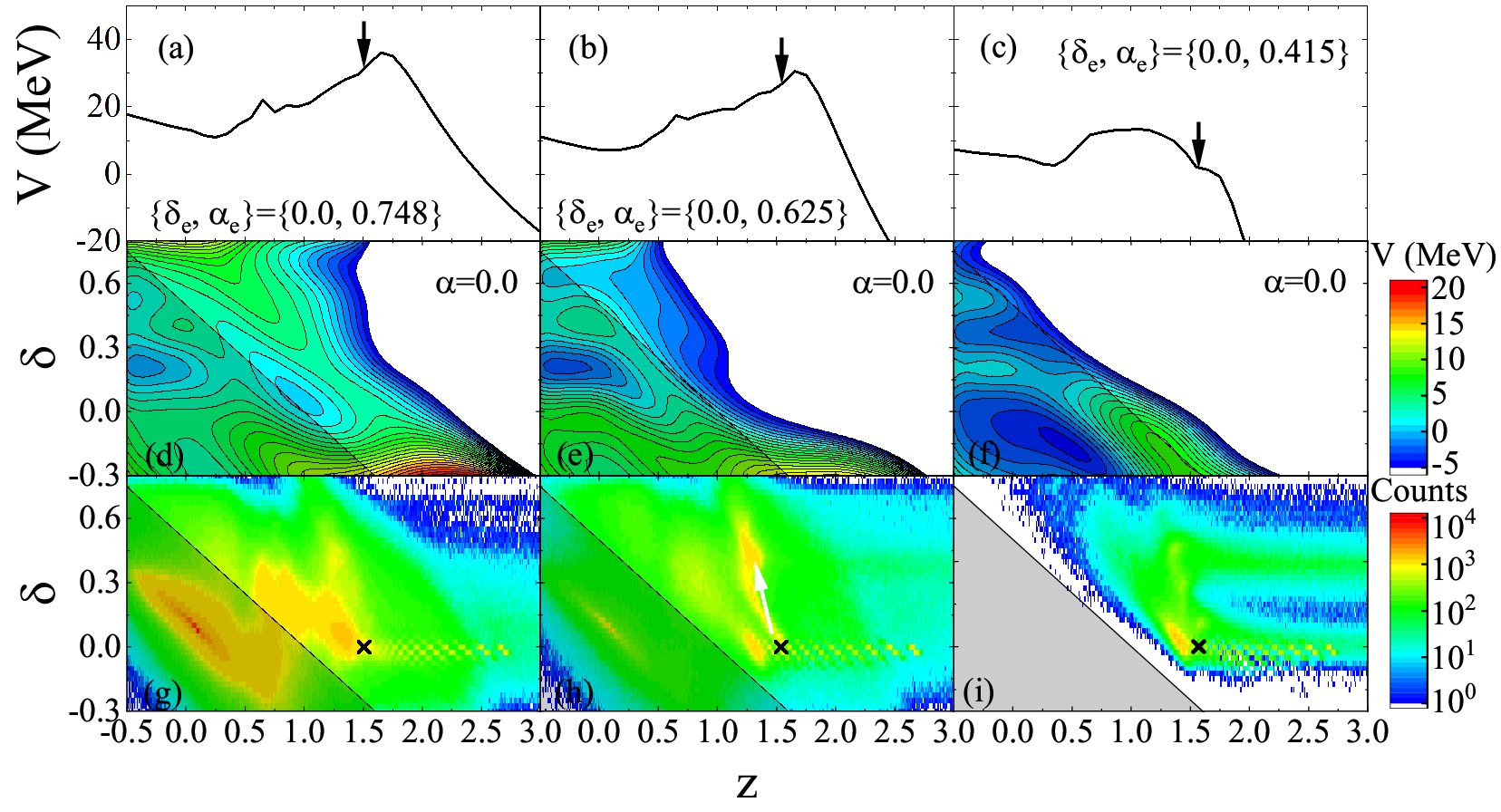}
\caption{One-dimensional barriers for the degree of entrance channel mass asymmetry $\alpha_\text{e}$ and deformation $\delta_\text{e}$ in the collision systems of (a) $^{30}$Si, (b) $^{48}$Ti, and (c) $^{86}$Kr with $^{208}$Pb. Black arrows indicate the contact point (a) $z=1.51$, (b) $z=1.54$, (c) $z=1.57$.
$z-\delta$ potential energy surface for the degree of the compound nucleus mass asymmetry $\alpha=0.0$ in the collision systems of (d) $^{30}$Si, (e) $^{48}$Ti, and (f) $^{86}$Kr with $^{208}$Pb. Panels (g) to (i) are trajectory distributions in the collision systems of $^{30}$Si, $^{48}$Ti, and $^{86}$Kr with $^{208}$Pb at $E_\text{c.m.}/V_\text{bass}=1.05$, $E_\text{c.m.}/V_\text{bass}=1.173$, and $E_\text{c.m.}/V_\text{bass}=1.084$. Right triangles in panel (d) to (i) shows the fusion box on $z-\delta$ plane.
Cross symbols show the contact point. All panels are in the case of central collisions.
}\label{fig4}
\end{figure*}
\begin{figure}[t]
\centering
\includegraphics[keepaspectratio, width=\linewidth]{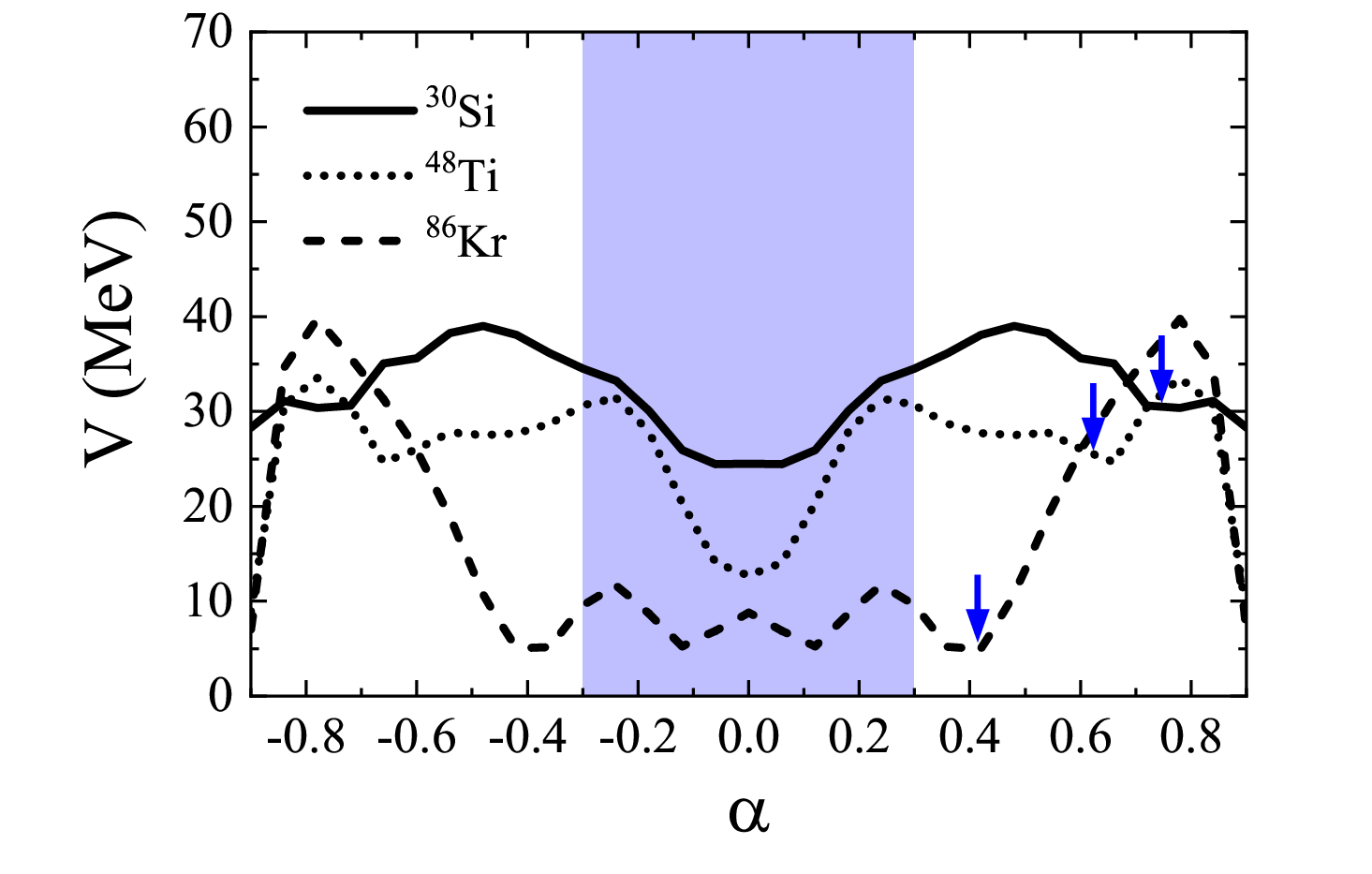}
\caption{Potential energy of the entrance channel as a function of mass asymmetry in the combination of fixing $^{208}$Pb as the target nucleus and selecting various projectile nuclei. The three blue arrows present the corresponding $\alpha_\text{e}$ for each system. The blue region indicates the fusion box for $\alpha$ defined in the text.
}\label{fig44}
\end{figure}
\begin{figure}[t]
\centering
\includegraphics[keepaspectratio, width=\linewidth]{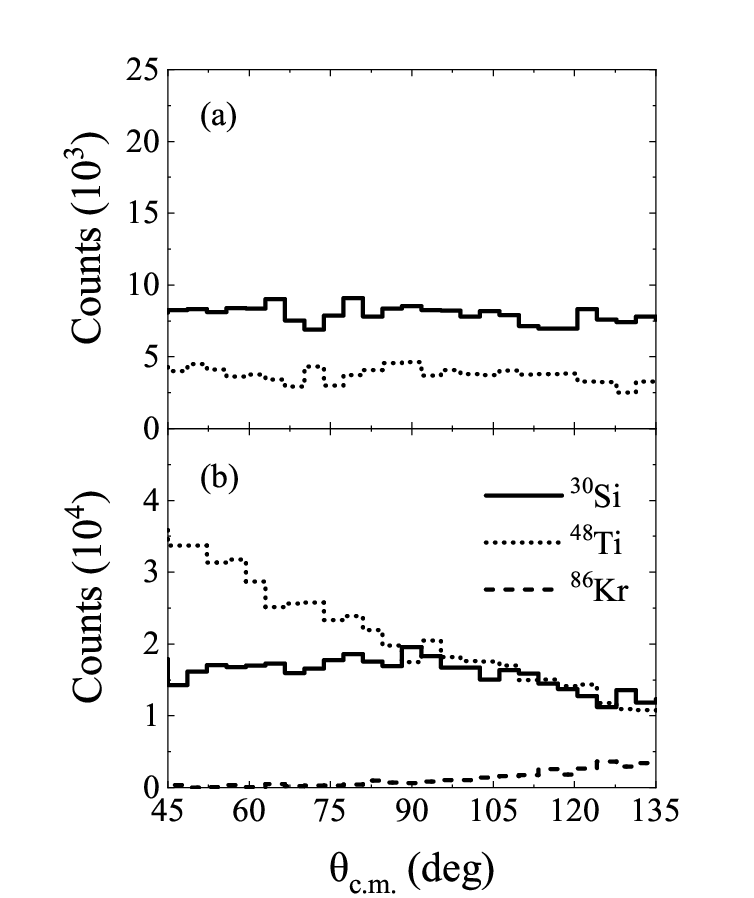}
\caption{Angular distributions of (a) FF component and (b) QF component selected for the different projectile nuclei using $^{208}$Pb as the target nucleus. The $M_\text{R}$ and angular ranges used correspond to a black rectangle in Fig.~\ref{fig3}(c)(i)(o).
}\label{fig5}
\end{figure}
In this section, we determine the appropriate value of the start time $t_0$ of the neck expansion by reproducing the experimental MAD and $M_\text{R}$ distribution using several entrance channels.
We prepare the reaction system having the different initial mass ratio $M_e^0 =m_P/(m_P + m_T)$ or $M_e^0 =m_T/(m_P + m_T)$, here $m_P , m_T$ are the projectile and the target masses. As the target we set $^{208}$Pb and as the projectile we choose $^{30}$Si, $^{48}$Ti and $^{86}$Kr. This analysis is also the extension study presented in the previous work \cite{amano2023dynamical}.

Figure~\ref{fig3} shows MADs and $M_\text{R}$ distributions depending on $t_0$.
The calculations with $t_0=0$ zs  shown in the left column correspond to the ``contact relaxation'' of the neck degree of freedom and the calculations with $t_0=5$ zs and $10$ zs in the middle and the right column corresponds to the ``delayed relaxation'' as explained in Ref. \cite{amano2023dynamical}.

The $M_\text{R}$ distribution of $^{30}$Si + $^{208}$Pb is dominated by the mass-symmetric fission for any value of $t_0$ as shown in Fig.~\ref{fig3}(d)-(f).
Among them, the calculation result using $t_{0}=10$ zs shows the sharper mass-symmetric fission (Fig.~\ref{fig3}(f)) and is in good agreement with the experimental data. Furthermore, in the calculated MAD using $t_{0}=10$ zs (Fig.~\ref{fig3}(c)), there is no correlation between mass and angle as shown in the experimental MAD \cite{williams2013evolution}.
In this reaction system, as shown in Fig.~\ref{fig4}(a), the one-dimensional entrance barrier has a large gap (about 30 MeV) against the fused system around $z=0$. 
Though the rapid expansion of the neck radius causes the eﬀect of the inner barrier for fusion hindrance \cite{amano2023dynamical}, the mass transfer does not change dramatically in this system. 
This is because the large gap mentioned above weakens the effect of the inner barrier and does not significantly suppress the mass equilibration related to fusion. 
It is noted that even if the effect of the inner barrier is weakened, the strong mass-angle correlation in Fig.~\ref{fig3}(a) appearing the predominant QF process implys the enough working of the fusion hindrance.

In $^{48}$Ti + $^{208}$Pb of heavier collision system (Fig.~\ref{fig3}(g)-(l)), the dominant fission mode changes dramatically from mass-asymmetric fission to mass-symmetric fission as $t_0$ increases. The results in Fig.~\ref{fig3}(i) and Fig.~\ref{fig3}(l) for $t_0=10$ zs show good agreement with the experimental results \cite{du2013mapping}.
The calculation result of Fig.~\ref{fig3}(l) shows sufficient progress of mass equilibration in contrast to the calculation result shown in Fig.~\ref{fig3}(j).
This can be seen also from the strong correlation between mass and angle in Fig.~\ref{fig3}(i).
Experimentally, the strong mass-angle correlation have been reported as evidence that QF events are dominant \cite{du2013mapping}.
The reason causing the mass-asymmetric fission in Fig.~\ref{fig3}(j) is due to the expansion of the neck at the initial stage of contact as described above.
As shown in Fig.~\ref{fig4}(b), the energy gap between the entrance barrier and the system around $z=0$ is about 20 MeV, and the fusion hindrance due to the neck expansion is effectively worked in comparison with the $^{30}$Si + $^{208}$Pb case. 
The fusion hindrance due to the neck expansion is considered there is the entrance channel dependence.

In $^{86}$Kr + $^{208}$Pb of even heavier collision system, the mass-asymmetric fission is dominant no matter when the significant expansion of the neck radius starts after contact. In particular, the calculation results using $t_{0}=5.0$ zs and $t_{0}=10$ zs are almost the same, and the calculated $M_\text{R}$ reproduced the experimental feature \cite{mg2002fusion} of $M_\text{R}$.
The $M_\text{R}$ in Fig.~\ref{fig3}(q) and Fig.~\ref{fig3}(r) populates close to the initial mass ratio $M_\text{R}^{0}$. This implies the dominant process is DIC events.
In the reactions dominated by DIC, the effects of the energy damping mode due to the strong Coulomb repulsion rather than the effects of the rapid expansion for the neck radius is significantly effective.
In the $^{86}$Kr + $^{208}$Pb collision system, the projectile-like nucleus shows backward scattering due to the kinetic energy being strongly damped (See Fig.~\ref{fig3}(n) and Fig.~\ref{fig3}(o)).
In Fig.~\ref{fig3}(m) and Fig.~\ref{fig3}(p), comparing with the case for $t_{0}=5.0$ and $10$ zs, massive nucleon transfer occurs through the neck immediately after contact because the neck radius rapidly expands before the two nuclei separate due to the strong Coulomb repulsion.

From the comparison of the experimental results and the calculated results, we inferred the ``delayed relaxation'' \cite{amano2023dynamical} of neck is preferable and use $t_{0}=10$ zs as the appropriate value in this paper.

In order to investigate further the MADs and $M_\text{R}$ distribution of the three types of  reactions, we compare the features of one-dimensional entrance barrier for each entrance channel in Figs.~\ref{fig4}(a)-(c).
As the degree of the entrance channel mass asymmetry $\alpha_e$ is smaller, the contact point shifts from the inside (small $z$) to the outside (large $z$) of the ridge at the contact barrier .
This simple shift for the contact point hinders the fusion of the system. This fusion hindrance has been reported in Refs. \cite{shen2009analysis,aritomo2006fusion}.
Figures~\ref{fig4}(d)-(f) show the potential energy surfaces on the $z-\delta$ plane at $\alpha=0.0$ corresponding to the compound nucleus mass in the different collision systems. The inner barrier significantly grows in the process of the degree of mass asymmetry relaxing from $\alpha_\text{e}$ to $\alpha=0.0$, as $\alpha_\text{e}$ of the collision system is smaller.

The feature of the potential energy surface closely relates to the behavior of the trajectory.
Figures~\ref{fig4}(g)-(i) show the trace of each trajectory on the $z-\delta$ plane in the case of central collisions. The fusion event is selected by whether a trajectory invades the fusion box $\lbrace |\alpha|<0.3, \delta<-0.5z+0.5) \rbrace$ or not as given in Refs. \cite{ARITOMO20043,PhysRevC.85.044614}.
The fusion box is indicated by the shaded triangles in left side corner of each panel.
In the collision system of $^{30}$Si + $^{208}$Pb, many trajectories invades the fusion box. The trajectories trapped into the pocket around \{$z$, $\delta$\} = \{0.0, 0.15\} (See Fig.~\ref{fig4}(d)) move to the direction of fission due to the fluctuation. We distinguish the FF and QF processes by analyzing whether the trajectory enters the fusion box or not. Consequently, FF is the main contribution in the collision system of $^{30}$Si + $^{208}$Pb.
Fewer trajectories in the case of $^{48}$Ti + $^{208}$Pb invade the fusion box in comparison with the collision system of $^{30}$Si + $^{208}$Pb.
On the contrary, the upward flow of trajectories (indicated by the white arrow) can be seen from the contact point into the direction of the large $\delta$ as if the trajectories are hindered from invading the fusion box.
In heavier collision system of $^{86}$Kr + $^{208}$Pb, some trajectories overcome the contact barrier by the effects of fluctuations, even if the contact point is at the outside against the ridge of the barrier.
However, all trajectories are hindered from invading the fusion box after contact by the developing inner barrier as the degree of mass asymmetry is relaxed toward $\alpha=0.0$.

The values of the initial mass asymmetry are quite different for these three reactions as shown in Fig.~\ref{fig4}. Therefore, we also mention the potential energy of the entrance channel from the perspective of the mass asymmetry. 
Figure~\ref{fig44} represents the potential energy of the entrance channel as a function of mass asymmetry $\alpha$. 
From the potential energy of the entrance channel in the case of $^{30}$Si, the trajectory is sucked into the fusion region (blue region) from $\alpha_\text{e}=0.748$ by the fluctuation effects after it overcomes the hump near $\alpha=0.5$. 
Note that the potential energy changes depending on the distance $z$ and the deformation $\delta$.

The potential energy in the case of the heavier $^{48}$Ti has two features. 
First, the gradual uphill slope of potential energy is confirmed from the point $ \alpha_\text{e}=0.625$ toward the fusion region. 
This uphill slope gently prevents the trajectory from entering the fusion region. 
Second, in contrast to the case of $^{30}$Si, the uphill slope (hump) toward the mass symmetry region is relatively flat. 
From this feature, the trajectory that cannot cross the top stated around $\alpha=0.25$ due to the above mentioned gentle fusion hindrance. As a result, the trajectory tends to settle down around $\alpha=0.4$ ($M_\text{R}=0.7$), and QF becomes dominant.

In even heavier $^{86}$Kr, there is a wall from $ \alpha_\text{e}=0.415$ toward the fusion region, but it can be easily overcome by fluctuations. Consequently, trajectories invade into the fusion box for $\alpha$. 
However, if we see in the $\delta$ space as shown in Fig.~\ref{fig4}(i), it is found that trajectories do not enter into the fusion box. Hence, the mass symmetry events mean QF, and DIC components are dominant outside the mass symmetric region.
It has already been reported that the formation of the compound nucleus (fusion event) is not related necessarily to the mass symmetric fission in Refs.~\cite{ARITOMO20043,PhysRevC.85.044614}. 
This tendency is particularly evident in the synthesis of heavier superheavy nuclei from both understandings of  Figs.~\ref{fig3}(f)(l)(r) and Figs.~\ref{fig4}(g)(h)(i).

Next, we investigate the angular distribution of FF and QF components in the different channels.
The calculation results are shown in Fig.~\ref{fig5} excluding the quasielastic scattering events.
Calculations were performed in both $M_\text{R}$ and $\theta_\text{c.m.}$ ranges $\lbrace0.19 \le M_\text{R} \le 0.81$, $45^\circ<\theta_\text{c.m.}<135^\circ\rbrace$, $\lbrace0.29 \le M_\text{R} \le 0.71$, $45^\circ<\theta_\text{c.m.}<135^\circ\rbrace$, and $\lbrace0.34 \le M_\text{R} \le 0.66$, $45^\circ<\theta_\text{c.m.}<135^\circ\rbrace$, which correspond to the black rectangles in Fig.~\ref{fig3}(c) and Fig.~\ref{fig3}(i), and Fig.~\ref{fig3}(o), respectively. 
FF events can not be seen in the collision system of $^{86}$Kr + $^{208}$Pb. There is no anisotropy to the scattering angles of FF (See Fig.~\ref{fig5}(a)) because the sticking time is long (several rotations before fission) and the entrance memory disappears.
In contrast with FF, the scattering angles of QF have a different anisotropy by each entrance channel (See Fig.~\ref{fig5}(b)). The anisotropy to the scattering angle of QF is weak in the collision system of $^{30}$Si + $^{208}$Pb. This characteristic is considered to be the contribution of slow QF (SQF) included in QF events.
In the case of SQF, while the projectile- and target-like nuclei rotate several times sticking together, the entrance memory gradually forgets by progressing the equilibration.
There is fast QF (FQF) which is another component in QF.
In FQF, the projectile- and target-like nuclei often split within half a rotation.

From Fig.~\ref{fig5}(b), the forward scattering is dominant in the collision system of $^{48}$Ti + $^{208}$Pb. This is because the projectile-like nucleus splits from target-like ones around half a rotation, which is the evidence that a lot of FQF events are included.
Looking toward QF component of $^{86}$Kr + $^{208}$Pb system, the backward scattering is dominant which means that the mass evolution toward $M_R=0.5$ allows only for low angular momentum, where the effects of $V_\text{rot}$ is weak. A similar characteristic at low angular momentum (small impact parameter) has been reported within TDHF calculations \cite{wakhle2014interplay,wakhle2015comparing,umar2015time}.

%
%
\subsection{Mass drift and the nuclear orientation}
\label {sec3B}
\begin{figure*}[t]
\centering
\includegraphics[keepaspectratio, width=\linewidth]{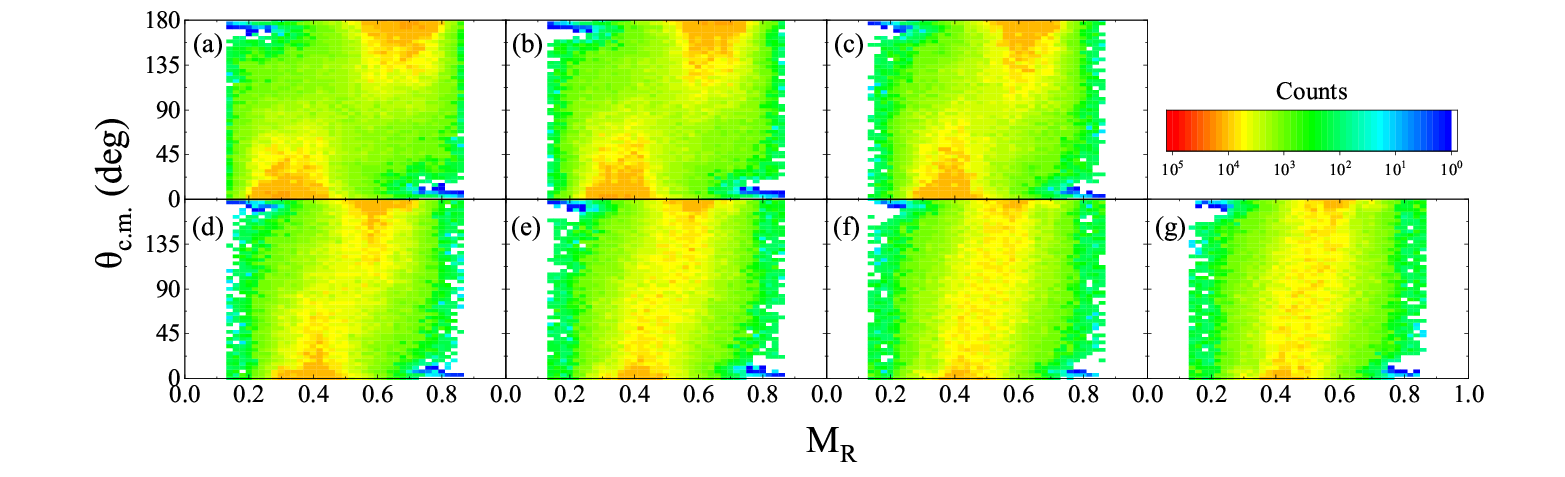}
\caption{Seven panels (a) to (g) show calculated MADs for seven nuclear orientations (0$^\circ$, 15$^\circ$, 30$^\circ$, 45$^\circ$, 60$^\circ$, 75$^\circ$, 90$^\circ$) of the deformed nucleus $^{232}$Th. The calculation data are for the $^{32}$S + $^{232}$Th system at $E_\text{c.m.}/V_\text{bass}=1.108$. The MADs show a transition from mass-asymmetric associated with the orientation angle 0$^\circ$ (tip collision) to mass-symmetric associated with the orientation angle 90$^\circ$ (side collision).
}\label{fig6}
\end{figure*}
\begin{figure}[t]
\centering
\includegraphics[keepaspectratio, width=\linewidth]{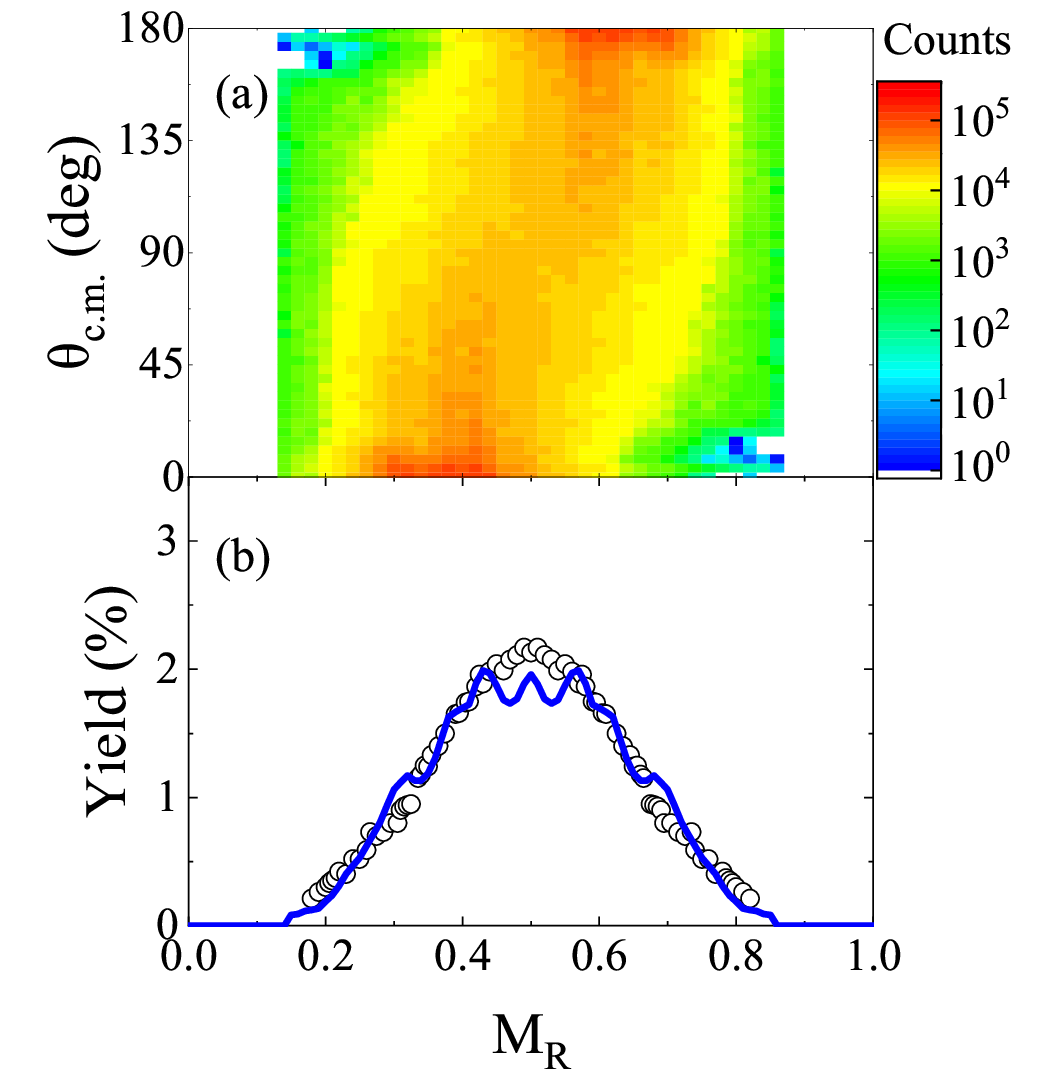}
\caption{(a) Mass angle distribution and (b) $M_\text{R}$ distribution calculated for $^{32}$S + $^{232}$Th at $E_\text{c.m.}/V_\text{bass}=1.108$. Calculations take into account the nuclear orientations of the deformed nucleus $^{232}$Th. Open circles are the experimental data shown in \cite{galkina2021investigating}
}\label{fig7}
\end{figure}
\begin{figure}[t]
\centering
\includegraphics[keepaspectratio, width=\linewidth]{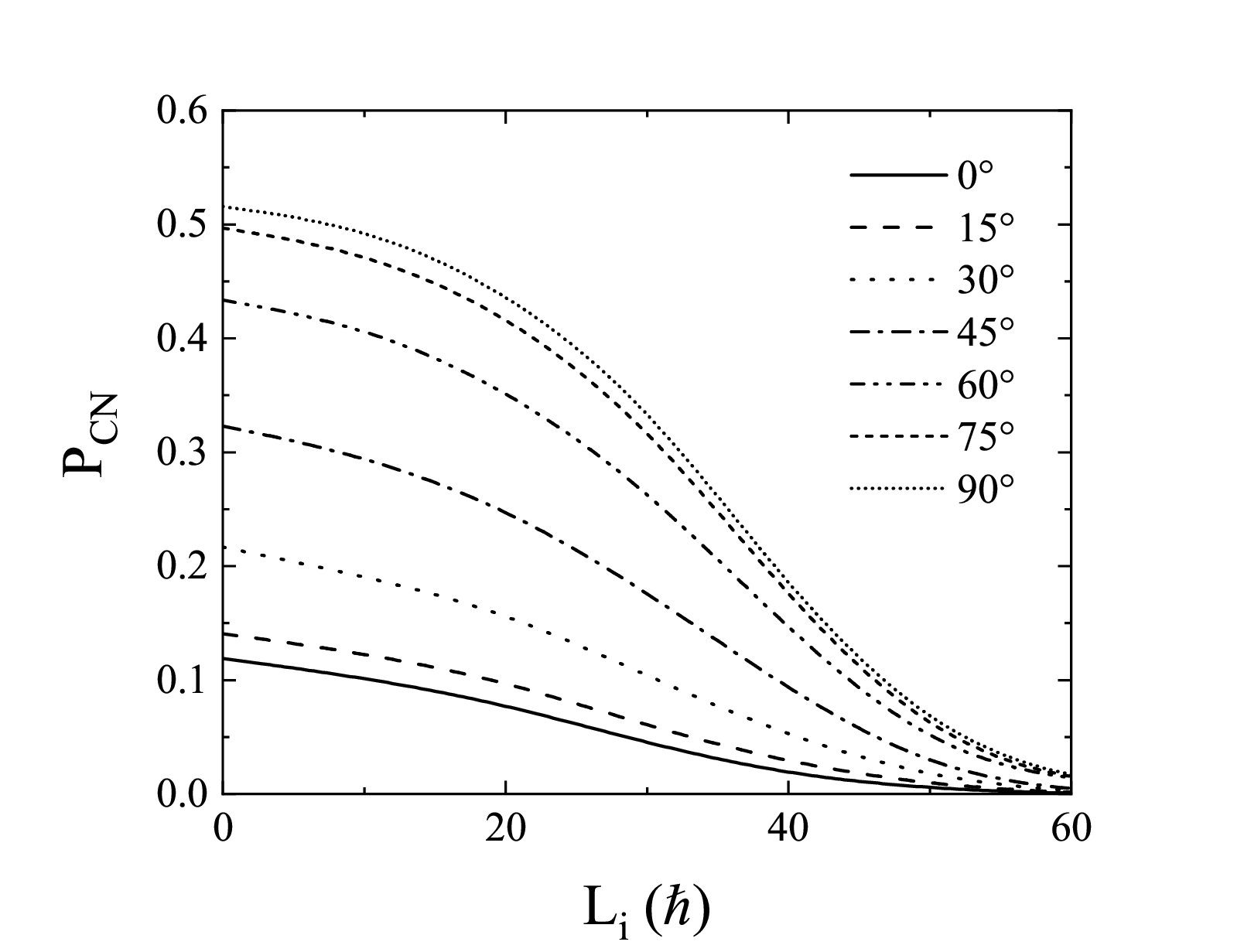}
\caption{Fusion probabilities for seven nuclear orientations (0$^\circ$, 15$^\circ$, 30$^\circ$, 45$^\circ$, 60$^\circ$, 75$^\circ$, 90$^\circ$) of the deformed nucleus $^{232}$Th plotted as a function of the orbital angular momentum $L_i$.
}\label{fig8}
\end{figure}
\begin{figure}[t]
\centering
\includegraphics[keepaspectratio, width=\linewidth]{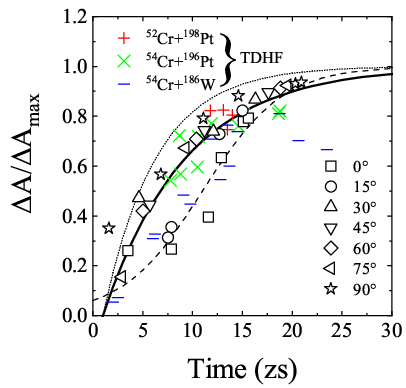}
\caption{Mass drift $\Delta A/\Delta A_\text{max}$ for the different nuclear orientation of the deformed nucleus $^{232}$Th as a function of the mean sticking time.
The symbols are data given by Table \ref{table2}, together with results of TDHF calculations obtained in Refs. \cite{tanaka2021mass,vo2018microscopic,simenel2020timescales,hammerton2015reduced}.
Solid and dotted curves are given with eq.~(\ref{eq_massdrift1}) using the parameter for the new scenario $\tau=8.3$ zs and the standard parameter $\tau=5.3$ zs \cite{PhysRevC.36.115}, respectively.
Dashed curve is also the new mass drift function given with eq.~(\ref{drift3}), assuming a Fermi-type mass drift.
}\label{fig9}
\end{figure}
\begin{figure}[t]
\centering
\includegraphics[keepaspectratio, width=\linewidth]{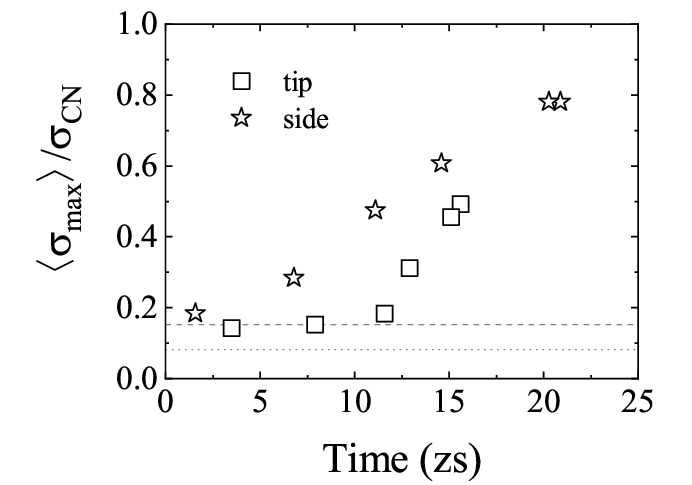}
\caption{Ratio of the mean maximum neck cross-sectional area $\langle\sigma_\text{max}\rangle$ (See text for the definition of $\langle\sigma_\text{max}\rangle$) to compound nucleus cross-sectional area $\sigma_\text{CN}$ as a function of the mean sticking time. The difinition of $\langle\sigma_\text{max}\rangle$ Open rectangles and stars are the calculated data for the tip collision case (0$^\circ$) and the side collision case (90$^\circ$), respectively. Horizontal dotted and dashed lines correspond to the neck cross-sectional area where the first mass drift begins in the tip collision and the side collision, respectively.
}\label{fig10}
\end{figure}

The mass drift between the reaction partner in FF and QF process has been studied in detail by Shen \textit{et~al}. \cite{PhysRevC.36.115}, and the systematics for the drift speed toward the direction of mass symmetry has discussed also. Here, we investigate the effect on the mass drift when the orientation of the reaction partner is taken into account. We study the deviation from the above mentioned systematics \cite{PhysRevC.36.115} and discuss the mass drift mode in QF process. In the following, the effect on our calculation taking account of the orientation of the deformed target in the collision of $^{32}$S + $^{232}$Th at $E_\text{c.m.}/V_\text{bass}=1.108$ is presented.
Here, the Bass barrier energy $V_\text{bass}$ in the $^{32}$S + $^{232}$Th is 158.55 MeV \cite{bass1980nuclear}.
The calculations are performed in the range $0.15<\text{M}_\text{R}<0.85$ (to simply eliminate quasielastic events).

Figure~\ref{fig6} shows the MADs of $^{32}$S~+~$^{232}$Th at $E_\text{c.m.}/V_\text{bass}=1.108$ in seven orientation angles of the deformed $^{232}$Th nucleus.
In Fig.~\ref{fig6}(a), we can see that the mass-asymmetric fission is dominant and the mass evolution is restricted to producing fragments with $M_{R}=0.3, 0.7$.
This restriction in the mass evolution is due to the narrow neck cross-sectional area rather than the effect of the shell structure, because the interaction energy is enough high above the barrier. The shell effect generally works for the subbarrier reactions.
The projectile and the target nucleus stick together in more than a half rotation, and as the orientation angle of $^{232}$Th becomes large, the mass equilibration progresses in the meantime.
In contrast with the tip collision case, although the MAD for the side collision (See Fig.~\ref{fig6}(g)) is dominated by the mass-symmetric fission, a significant correlation between mass and angle can still be seen.
This result implies that QF events are dominant even for the side collision case.
Namely, even if the mass-symmetric fission events are dominant, many of their trajectories do not path through the compound nucleus area.

The integrated MAD and $M_\text{R}$ distribution over the nuclear orientations of the deformed nucleus $^{232}$Th are shown in Fig.~\ref{fig7}.
A strong correlation between mass and angle can be seen in Fig.~\ref{fig7}(a) and the experimental trend \cite{du2013mapping} is well reproduced.
Figure~\ref{fig7}(b) shows MAD projected onto the $M_\text{R}$ axis and the solid curve which gives a good agreement with the experimental trend \cite{galkina2021investigating}. 
The experimental results are inferred to be a mix of fission products in all nuclear orientation angles because the incident energy is sufficiently high. 
The calculation result in Fig.~\ref{fig7}(b) is obtained by weighting the data for all orientation angles equally. 
In this treatment, particularly the calculation data around $M_\text{R}=0.5$ seems to be a little different from the experimental data. 
In Refs. \cite{hinde1996conclusive,wakhle2014interplay}, the relative weighting was performed on the orientation angles at the collision. Therefore, we consider that the weighting that increases the contribution of fission products is needed as the orientation angle changes from 0$^\circ$ to 90$^\circ$, against the present calculation data assuming to mix all orientations equally. 
More quantitative analysis of $M_\text{R}$ is a future task.

It is reported that the orbital angular momentum affects not only rotation angle but also mass equilibration in TDHF calculations \cite{wakhle2014interplay,umar2015time,umar2016fusion,godbey2019deformed}.
Sufficient mass equilibration is considered to lead to the fusion of the collision system.
Here, we estimate the fusion probability $P_\text{CN}$ dependent on the orientation angle of $^{232}$Th nucleus.
The fusion probability $P_\text{CN}$ is shown in Fig.~\ref{fig8} as a function of the orbital angular momentum $L_i$.
The calculation is performed with seven orientation angles of $^{232}$Th target.
As the larger nuclear orientation angle at the collision, $P_\text{CN}$ becomes higher. $P_\text{CN}$ is the highest for the side orientation (90$^\circ$) through the range of $L_i$. The fusion events gently decrease toward high $L_i$, because the trajectories cannot invade the fusion pocket due to the increase of the fusion barrier composed by increasing the rotational energy $V_\text{rot}$ \cite{amano2023dynamical}.
The realization of the equilibrium system (fusion) is correlated with the evolution of mass and rotation angle in the MAD.
However, in superheavy mass region, it has been confirmed from both theoretical and experimental analysis that the fusion hindrance occurs even in the range of $A_\text{CN}/2\pm20$ u \cite{ARITOMO20043, PhysRevC.85.044614, PhysRevC.94.054613}. Therefore, the mass equilibration does not necessarily lead directly to fusion.
The trajectories in this region do not necessarily cross the fusion area.

Next, we investigate the correlation between mass evolution and sticking time to see the feature of the mass drift toward the mass equilibration in detail.
The degree of mass drift toward symmetry $\Delta A/\Delta A_\text{max}$ which is the function of time is given by
\begin{eqnarray}
\frac{\Delta A}{\Delta A_\text{max}}=\frac{A_t-\langle A \rangle}{A_t-A_s},
\label{eq_massdrift1}
\end{eqnarray}
where $\langle A \rangle$ represents the mean mass of the target-like nucleus at each angular momentum. $A_s$ represents the symmetric mass expressed as
\begin{eqnarray}
A_s=\frac{1}{2}(A_p+A_t),
\label{eq_massdrift2}
\end{eqnarray}
where $A_p$ and $A_t$ are the projectile and target mass.
For the evaluation of the mean sticking time $\langle t_{s} \rangle$, it is necessary to consider the time when the neck forms sufficiently after contact and the time of neck shrinking prior to scission \cite{PhysRevC.36.115}.
We estimate $\langle t_{s} \rangle$ as follows: First, we calculate the initial neck radius $R_\text{n}^\text{in}~(=\!\!R_\text{n})$ when even one nucleon has transferred between two nuclei after the projectile and target nucleus come into contact. 
By two-center parameterization, the neck radius $R_\text{n}$ is obtained using eq.~(7) in Ref. \cite{amano2023nuclear}, 
and by performing many trajectory calculations, various $R_\text{n}^\text{in}$ values are estimated for each trajectory. 
By averaging these values, the mean initial neck radius $\langle R_\text{n}^\text{in} \rangle$ is estimated. 
Secondly, we set $t=0$ zs at the point when the neck radius grows to $\langle R_\text{n}^\text{in}\rangle$.
Finally,  $t_s$ is defined as the time from $t=0$ zs to the time when the neck radius shrunk again to the previous $\langle R_\text{n}^\text{in}\rangle$ prior to scission.
The values of $\langle R_\text{n}^\text{in}\rangle$ which the ﬁrst nucleon transfer begins is shown for each orientation angle in Table \ref{table1}. 
%
%
\renewcommand{\arraystretch}{1.2}
\begin{table}[h]
  \caption{Orientation angle of the deformed $^{232}$Th nucleus and the mean initial neck radius $\langle R_\text{n}^\text{in}\rangle$ when the first mass drift begins after contact.}
  \label{table1}
  \centering
  \begin{tabular}{cccccccc} \hline
   Orientation angle (deg) & 0 & 15 & 30 & 45 & 60 & 75 & 90 \\ \hline
   $\langle R_\text{n}^\text{in} \rangle$ (fm) & 2.2 & 2.4 & 2.6 & 2.8 & 2.9 & 2.9 & 3.0 \\ \hline
 \end{tabular}
\end{table}
\renewcommand{\arraystretch}{1}
The increase in $\langle R_\text{n}^\text{in}\rangle$ with the increase of the orientation angle is mainly caused by the difference of the geometrical contact area between two nuclei.
The information of both $\langle A \rangle$ and $\langle t_s \rangle$ obtained from our calculations is shown in Table \ref{table2}.
%
%
\begin{table}[t]
  \caption{Information of orientation angles for the deformed $^{232}$Th nucleus in the $^{32}$S + $^{232}$Th system, mean angular momentum $\langle L \rangle$, mean target-like mass $\langle A \rangle$, mass drift $\Delta A/\Delta A_\text{max}$, mean sticking time $\langle t_s \rangle$.
}
  \label{table2}
  \centering
  \begin{tabular}{ccccccc}
    \toprule
    Orientation angle & $\langle L \rangle$ & \multirow{2}{*}{$\langle A \rangle$} & \multirow{2}{*}{$\frac{\Delta A}{\Delta A_\text{max}}$} & $\langle t_s \rangle$
\\[2pt]
      (deg) & ($\hbar$) &  &  & (zs) \\
    \midrule
    0 &  2 & 153 & 0.79 & 15.6 \\
       & 15 & 154 & 0.78 & 15.1 \\
       & 44 & 169 & 0.63 & 12.9 \\
       & 77 & 192 & 0.40 & 11.6 \\
       & 96 & 205 & 0.27 &  7.9 \\
      & 102 & 206 & 0.26 & 3.5 \\
    15 & 12 & 150 & 0.82 & 15.0 \\
         & 39 & 159 & 0.73 & 12.7 \\
         & 95 & 197 & 0.35 & 7.9 \\
       & 100 & 201 & 0.31 & 7.5 \\
    30   & 4 & 145 & 0.87 & 16.3 \\
         & 45 & 158 & 0.74 & 12.1 \\
       & 103 & 185 & 0.47 & 4.6  \\
    45   & 7 & 142 & 0.90 & 17.6 \\
         & 50 & 158 & 0.74 & 11.2 \\
         & 96 & 187 & 0.45 & 5.7  \\
    60 & 14 & 141 & 0.91 & 18.8 \\
         & 57 & 161 & 0.71 & 10.4 \\
         & 95 & 190 & 0.42 & 5.0   \\
    75 & 16 & 140 & 0.92 & 19.5 \\
         & 62 & 165 & 0.67 & 9.3   \\
         & 98 & 216 & 0.16 & 2.9  \\
    90 &  5 & 138 & 0.94 & 20.9 \\
         & 11 & 139 & 0.93 & 20.3 \\
         & 36 & 144 & 0.88 & 14.6 \\
         & 49 & 153 & 0.79 & 11.1 \\
         & 73 & 175 & 0.57 & 6.8 \\
       & 100 & 197 & 0.35 & 1.6  \\
    \bottomrule
 \end{tabular}
\end{table}

Figure~\ref{fig9} shows $\Delta A/\Delta A_\text{max}$ for each orientation angle as a function of the mean sticking time corresponding to the induced mean angular momentum $\langle L \rangle$ indicating in Table \ref{table2}.
The dotted curve shown in Fig.~\ref{fig9} presents the drift mode \cite{PhysRevC.36.115} toward the mass-symmetry as the overdamped motion following
\begin{eqnarray}
\frac{\Delta A}{\Delta A_\text{max}}=1-\text{exp}\left\lbrack \frac{-(t-t_d)}{\tau}\right\rbrack,
\label{drift_shen}
\end{eqnarray}
where $\tau$ was determined as 5.3 zs based on the correlation between $\langle A \rangle$ and $\langle t_s \rangle$, obtained by $^{238}$U induced reactions \cite{PhysRevC.36.115}. The delay time $t_d$ was determined $1$ zs considering both the time reaching a sufficient growth of neck to begin the mass drift after contact and the neck shrinking time to finish the mass drift prior to scission. These values have been used as the standard value \cite{PhysRevC.36.115,tanaka2021mass}.
Regarding the delay time, the TDHF calculations \cite{simenel2012influence} has reported that before a significant mass drift an initial delay of $\le2$ zs is required for $N/Z$ equilibrium.
In our previous research \cite{PhysRevC.106.024610}, we can confirm that the delay time is $\sim2$ zs in order to occur massive nucleon transfer and during the delay time the square of neck radius increases up to $4-6$ fm$^2$. However, following Ref. \cite{PhysRevC.36.115} we use $t_{d}=1$ zs as the standard value in this paper.

As can be seen in Fig.~\ref{fig9}, there is a deviation between our calculation data and the standard mass drift curve (dotted curve). The calculation data \cite{tanaka2021mass,vo2018microscopic,simenel2020timescales,hammerton2015reduced} with TDHF method also shows a similar deviation.
These deviation implies that the mass evolution toward mass symmetry is slower than the mass evolution estimated by the standard quasifission analysis \cite{PhysRevC.36.115}.
At the intermediate angular momentum (40 $\le$ $L$ $\le$ 80), all calculated mass drift data deviate more from the standard mass drift curve in comparison with the data at other angular momentum ranges.
From the experimental analysis, it has been reported that the deviation at intermediate angular momentum comes from the strong role of fluctuations contributing significantly to the mass evolution from FQF process at intermediate angular momentum to DIC at high angular momentum \cite{tanaka2021mass}.

The slow mass drift mode using $\tau=8.3$ zs is shown in the solid curve. The calculation data are consistent with the slow mass drift mode than the standard mass drift one.
However, the mass drift of the tip collision ($0^{\circ}$) still significantly deviates from the slow mass drift curve. The origin of this large deviation is discussed later.
In Ref. \cite{prasad2016exploring}, simulated $M_\text{R}$s assuming the slow mass drift have been reported to be consistent with the experimental $M_\text{R}$ distribution.

Before discussing why the mass drift for the tip collision does not follow a prescription using eq.~(\ref{drift_shen}), we suggest the new mass drift function to the tip collision.
For the tip collision, the mass evolution is noticeably slow at intermediate $L$ in contrast with the standard quasifission ansatz \cite{PhysRevC.36.115}.
The new mass drift function reflecting this characteristic mass evolution in the tip collision is shown by the dashed curve in Fig.~\ref{fig9}, which is given by
\begin{eqnarray}
\left(\frac{\Delta A}{\Delta A_\text{max}}\right)_\text{tip}=1-\frac{1}{1+\text{exp}\left\lbrack (t-t_h)/\tau_\text{s} \right\rbrack}.
\label{drift3}
\end{eqnarray}
This function involves two parameters, that is, a parameter $t_h$ when mass transfer reaches half of the maximum, and $\tau_{s}$ the slope of function. We use $t_h$ and $\tau_{s}$ as 11 zs and 4 zs, respectively.
The value of two parameters was determined considering the correlation between mass evolution and sticking time.
This function presents the restriction of mass evolution until mass transfer reaches half of the maximum as can be seen in Fig.~\ref{fig9}.
After mass transfer reaches half of the maximum, the mass equilibration rapidly progresses.

Lastly, we discuss the reason why the characteristic mass drift mode appears in the tip collision. The difference of the drift modes toward $M_\text{R}=0.5$ between the tip collision ($0^{\circ}$) and the side collision ($90^{\circ}$) comes from the maximum neck cross-sectional area during the process from contact to scission. 
The neck radius is varying from the contact stage to the end of fission process. The maximum neck radius $R_\text{n}^\text{max}$ is realized at a certain time in this process, and $\pi{R_\text{n}^\text{max}}^2$ is defined as the maximum cross-sectional area $\sigma_\text{max}$. Besides, over many trials of trajectory calculation, we estimate the mean maximum neck cross-sectional area $\langle\sigma_\text{max}\rangle$. 
Figure~\ref{fig10} shows the ratio of the mean maximum neck cross-sectional area $\langle\sigma_\text{max}\rangle$ to the full cross-sectional area for compound nucleus $\sigma_\text{CN}=\pi R_\text{CN}^2$ as a function of the mean sticking time.
$\langle\sigma_\text{max}\rangle$ for the side collision increase linearly as time passes.
In comparison with the side collision, the increase of $\langle\sigma_\text{max}\rangle$ for the tip collision is suppressed up to the middle sticking time (corresponding to intermediate $L$).
Because of this, the sufficient neck cross-sectional area for the nucleon transfer cannot be kept enough even at intermediate $L$. Therefore, in the tip collision, the mass evolution toward mass symmetry is restricted.

\section{Summary}
The start time $t_0$ relevant to the neck expansion has been discussed in the several entrance channels using the dynamical model based on a fluctuation-dissipation theorem.
The MADs obtained in our calculation were confirmed to categorize into three types in consistent with experiments.
Focusing on both fission fragment mass and scattering angle, we presumed the ``delayed relaxation'' of the neck is preferable in the comparison with the experimental results, and we determined $t_0=10$ zs as the appropriate value.
This implies the neck connecting the two nuclei does not immediately enlarge after contact.
In particular, the role of ``delayed relaxation'' of the neck is important for the origin of the strong correlation between mass and angle.
Snapshots in Ref. \cite{adam1968collision} which have presented to maintain dumbbell shape ($\epsilon=1.0$) for a certain period of time after contact are consistent with our calculation results.

We have performed the model calculation for the MAD and $M_\text{R}$ considering the orientations of the target nucleus.
The variation of the mass evolution and the rotation angle of fragments have clarified changing the orientation of the target nucleus from $0^{\circ}$ (the tip collision) to $90^{\circ}$ (the side collision).
The integrated MAD and $M_\text{R}$ distribution over the nuclear orientations are in good agreement with the feature of the experimental results.  However, further investigations are needed for more precise reproduction of the experimental $M_\text{R}$. 
More importantly, the strong mass-angle correlation was also reproduced.
It is found that the fusion probability for the side collision is highest for any orbital angular momentum.
%

As for the mass drift, we have presented the deviation of our model calculation from the standard mass drift curve obtained by the quasifission analysis of $^{238}$U-induced reactions \cite{PhysRevC.36.115}.
 This deviation implies that the mass evolution toward mass symmetry is slower than the standard mass drift mode.
For the side collision, the correlation between mass evolution and sticking time can be understood by the slow mass drift function using an exponential-type.
On the other hand, the mass drift data of the tip collision do not follow the slow mass drift function.
In the case of the tip collision, the characteristic mass equilibration is restricted for intermediate $L$, and we suggest a Fermi-type function as a new mass drift mode.
The reason for the different mass drift modes in the tip and side collision is coming from the different feature of the maximum neck cross-sectional area appeared in the process of mass evolution (from contact to scission).
In the tip collision, the restriction of the mass evolution in intermediate $L$ is caused by the insufficient neck evolutional area in intermediate $L$ in contrast with the side collision.


\begin{acknowledgments}
The Langevin calculation were performed using the cluster computer system (Kindai-VOSTOK) which is supported by JSPS KAKENHI Grant Number 20K04003 and Research funds for External Fund Introduction 2021 by Kindai University.
\end{acknowledgments}
%
%


\end{document}